\documentclass[prd,twocolumn,nofootinbib,notitlepage,10pt,aps]{revtex4-1}
\usepackage[utf8]{inputenc}
\usepackage{cancel}
\usepackage{xcolor}
\usepackage{latexsym}
\usepackage{amsmath}
\usepackage{amssymb}
\usepackage{bbm}
\usepackage{tensor}
\usepackage{natbib}

\usepackage{hyperref}
\hypersetup{
    colorlinks,
    linkcolor={violet},
    citecolor={teal},
    urlcolor={teal}
}
\usepackage{orcidlink}

\makeatletter
\newcommand{\subseconly}[1]{\expandafter\@secondoftwo\csname r@#1\endcsname}
\makeatother

\usepackage{ulem}
\usepackage{pdfsync}
\usepackage{epsfig}
\usepackage{epstopdf}

\usepackage{subfigure}
\usepackage{color}
\usepackage{comment}
\usepackage{slashed}
\usepackage{physics}

\def\beq{\begin{equation}}
\def\eeq{\end{equation}}
\def\baq{\begin{eqnarray}}
\def\eaq{\end{eqnarray}}

\newcommand{\be}{\begin{equation}} 
\newcommand{\ee}{\end{equation}}
\newcommand{\bea}{\begin{equation} \begin{aligned}}
\newcommand{\eea}{\end{aligned} \end{equation}}

\begin{document}

\title{Toward a gravitational network: Bridging metric-affine gravity \\ and no-scale supergravity}

\author{Ioannis D. Gialamas\orcidlink{0000-0002-2957-5276}}
\email{ioannis.gialamas@kbfi.ee}
\affiliation{Laboratory of High Energy and Computational Physics, NICPB, R\"avala 10, 10143 Tallinn, Estonia}

\author{Theodoros Katsoulas\orcidlink{0000-0003-4103-7937}}
\email{th.katsoulas@uoi.gr}
\affiliation{Physics Department, University of Ioannina, 45110, Ioannina, Greece}

\author{Kyriakos Tamvakis\orcidlink{0009-0007-7953-9816}}
\email{tamvakis@uoi.gr}
\affiliation{Physics Department, University of Ioannina, 45110, Ioannina, Greece}

\begin{abstract}
We consider a class of models in the framework of metric-affine gravity and establish their correspondence to the bosonic sector of a class of no-scale supergravity models. The excellent inflationary behavior of the gravitational models considered is carried over to the corresponding supergravity ones, thus, enriching the landscape of inflation-compatible models.
\end{abstract}

\maketitle

%

\section{Introduction}

Most of the present investigations in cosmology are based on the working assumption that the quantum aspects of gravitation can be ignored for energies below the Planck scale and, therefore, gravity can be treated classically, while in contrast the full quantum character of standard model (SM) particle interactions is considered within quantum field theory. The quantum interactions of gravitating matter fields introduce modifications to the standard general relativity action with cosmological implications that can lead to measurable imprints on inflation observational data~\cite{Planck:2018jri,BICEP:2021xfz,ACT:2025fju,ACT:2025tim}. Such modifications are nonminimal couplings of matter fields to curvature or higher powers of curvature invariants. 
In the light of these developments there is a revived interest on more general formulations of the theory of gravitation such as the {\textit{metric-affine}} formulation of gravity, where not only the metric $g_{ \mu\nu}$ but also the affine-connection $\tensor{\Gamma}{^\rho_\mu_\nu}$ is an independent variable. In this framework, apart from the curvature, gravity is also characterized by nonzero {\textit{torsion}} $\tensor{\cal{T}}{^\rho_\mu_\nu}=2\Gamma^{\rho}_{\,\,[\mu,\nu]}$ and {\textit{nonmetricity}} $Q_{\rho \mu\nu}=\nabla_{ \rho}g_{\mu\nu}$. A particular case of interest is the so-called {\textit{Einstein-Cartan}} gravity, where the nonmetricity vanishes. Consistent models in this framework should not contain higher than quadratic terms of the curvature and the torsion and should be characterized by a healthy spectrum, not containing any ghostlike states~\cite{BeltranJimenez:2019hrm,BeltranJimenez:2019acz,Percacci:2020ddy,Marzo:2021iok,Baldazzi:2021kaf,Barker:2024ydb,Dyer:2024kvo,Barker:2025xzd}. This narrows the possible actions to those containing at most quadratic terms of the Ricci curvature scalar ${\cal{R}}$ and the Holst invariant  $\tilde{\cal{R}} = \tensor{\epsilon}{_\mu^\nu^\rho^\sigma}\tensor{\mathcal{R}}{^\mu_\nu_\rho_\sigma} $, as well as nonminimal coupling terms of scalars\footnote{Derivative coupling terms to the Ricci tensors $(\partial_{ \mu}\phi)(\partial_{ \nu}\phi){\cal{R}}^{\mu\nu}$ are also allowed.}  $f(\phi){\cal{R}},\,g(\phi)\tilde{\cal{R}}$. It turns out that, since both ${\cal{R}}$ and $\tilde{\cal{R}}$ can be expressed in terms of metric quantities, there is an equivalent metric expression of the action of these models. In this action the components of torsion, having at most a linear derivative term will satisfy an algebraic equation of motion. The final equivalent metric action may contain extra degrees of freedom of gravitational origin, such as an axionlike pseudoscalar attributed to the presence of a Holst term. It should be noted that as a general rule for these models their inflationary behavior is intimately linked with gravity, additional fine tunings of potential parameters are absent and inflation arises in a more or less generic fashion.

In an entirely different context, with the SM at its center, quite a few years ago supergravity was considered as a framework for all interactions. The extreme smallness of the observed vacuum energy has led to a particular class of supergravity models characterized by a naturally vanishing cosmological constant. The simplest of these {\textit{no-scale supergravity models}}~\cite{Cremmer:1983bf,Ellis:1983sf,Ellis:1984bm} (see the Appendix~\ref{appendix}), characterized by two chiral superfields $T$ and $S$ and a K{\"a}hler potential ${\cal{K}}=-3\ln\left(T+\bar{T}-h(S,\bar{S})\right)$, for a particular choice of superpotential $W(S,T)$, has been shown to reproduce the Starobinsky model~\cite{Cecotti:1987sa,Ellis:2013xoa,Ellis:2013nxa,Lahanas:2015jwa} as well as its corrections~\cite{Antoniadis:2025pfa,Gialamas:2025ofz}. It turns out that there is a much wider class of correspondences between gravitational models resulting from the above metric-affine framework and the bosonic sector of no-scale supergravity models.

It should be stressed that the present article explores a broad class of no-scale supergravity models with excellent inflationary properties, fully compatible with current observational data. 
Although these models are constructed by adapting specific superpotentials inspired by the successful metric-affine gravitational models, the correspondence between the two frameworks is indirect and primarily phenomenological\footnote{Models in the metric-affine framework, set up in their equivalent metric formulation are matched with corresponding supergravity models. In many cases the matter content, entirely of geometric/gravitational origin, and their interactions are very specific, resulting in specific choices of the corresponding supergravity superpotentials. This procedure is a generalization of the matching of the Starobinsky model with supergravity for a specific choice of superpotential~\cite{Ellis:2013xoa}.}. 
Nevertheless, this interplay enriches the theoretical landscape of inflationary model building and highlights the versatility of the no-scale framework.

In section~\ref{framework}, we present the basic framework of metric-affine gravity and no-scale supergravity. In section~\ref{models}, we analyze the metric-affine models and their correspondence with the no-scale supergravity models, identifying suitable superpotentials. Finally, we summarize our findings and conclude in section~\ref{conclusions}. 

In what follows, we denote the Ricci scalar in standard metric gravity by $R\equiv R[g]$, while in metric-affine gravity by $\mathcal{R}\equiv \mathcal{R}[g,\Gamma]$. We also set the reduced Planck mass to unity, $M_P=1$.

\section{Framework}
\label{framework}

In this section, we review tools from metric-affine gravity and no-scale supergravity that will be used in the next section.
\\[0.2cm]
\textit{Metric-affine gravity}--- We consider models with actions, in the Einstein-Cartan framework (i.e. vanishing nonmetricity), of the general form
\begin{align}
\label{GENLANG}
\mathcal{S} = \int {\rm d}^4x\sqrt{-g} &\left[\frac{1}{2}f(\phi){\cal{R}}+\frac{1}{2}g(\phi)\tilde{\cal{R}}+ \frac{\gamma}{4}{\cal{R}}^2+\frac{\delta}{4}\tilde{\cal{R}}^2 \right. \nonumber
\\
& \hspace{0.2cm}+ \left.\frac{\epsilon}{2}\mathcal{R}\tilde{\mathcal{R}}-\frac12(\partial_\mu\phi)^2 -V(\phi)\right]\,.
\end{align}
This action can be set into a metric-equivalent form by expressing the curvature scalars as
\begin{align}
\mathcal{R} &=R+2\nabla_\mu \mathcal{T}^\mu-\frac{2}{3}\mathcal{T}_\mu \mathcal{T}^\mu+\frac{1}{24}\hat{\mathcal{T}}_\mu \hat{\mathcal{T}}^\mu +\frac12 {\tau}_{\mu\nu\rho}\tau^{\mu\nu\rho}\,, \nonumber
\\
\tilde{\mathcal{R}} &= -\nabla_\mu \hat{\mathcal{T}}^\mu +\frac23 \hat{\mathcal{T}}_\mu \mathcal{T}^\mu +\frac12 \epsilon^{\mu\nu\rho\sigma} \tau_{\lambda\mu\nu} \tensor{\tau}{^\lambda_\rho_\sigma}\,,
\end{align}
where $\nabla_\mu$ is defined in terms of the standard Levi-Civita connection. The torsion vectors are defined as $\mathcal{T}_\mu = \tensor{\mathcal{T}}{^\rho_\mu_\rho}$ and $\hat{\cal{T}}^{\mu}=\epsilon^{\mu\nu\rho\sigma}{\cal{T}}_{\nu\rho\sigma}$. The traceless tensorial part of the torsion enters the action quadratically, and its equation of motion leads to $\tau_{ \mu\nu\rho}=0$, meaning that it can be set to zero to begin with. In each of the cases to be considered in Sec.~\ref{models}, the equivalent metric action is obtained by integrating out the nondynamical degrees of freedom\footnote{For the models~\hyperref[modelA]{A} and~\hyperref[modelB]{B} considered in Sec.~\ref{models}, an equivalent formulation can be applied either by setting torsion to zero and allowing nonmetricity to remain nonzero, or by keeping both torsion and nonmetricity nonvanishing, leading to the same equivalent metric action. In models~\hyperref[modelC]{C} and~\hyperref[modelD]{D}, torsion is nonzero, while nonmetricity can arise by a projective transformation, still resulting in the same metric-equivalent theory. In the case of the locally Weyl-invariant models coupled to matter (model~\hyperref[modelE]{E}) the torsion is necessary, since the torsion vector is identified with the gauge vector of the underlying local gauge symmetry.}
\\[0.2cm]
\textit{No-scale supergravity}---This class of supergravity models, characterized by vanishing vacuum energy, is defined in terms of two chiral superfields described by a K{\"a}hler potential ${\cal{K}}=-3\ln(T+\overline{T}-h(S,\overline{S}))$. The function $h(S,\overline{S})$ for the simplest member of the class, namely the $SU(2,1)/SU(2)\times U(1)$ no-scale model is just $h(S,\overline{S})=|S|^2/3+\cdots$, where the dots signify higher powers of $|S|^2$ required for stability, while more general choices with $h_{S\overline{S}}(0)>0$ are possible. Introducing a superpotential $W(S,T)$ for the model, we obtain a nonvanishing scalar potential (see the Appendix~\ref{appendix}). In what follows we shall adopt a superpotential of the form 
\be
\label{eq:super_pot}
W(S,T)=SZ(T)\,,    
\ee
and consider the bosonic sector of the model in the limit $S\rightarrow 0$. This ansatz is analogous to the one employed in the derivation of the Starobinsky model~\cite{Starobinsky:1980te} in this framework~\cite{Ellis:2013nxa,Lahanas:2015jwa} with the specific choice $Z(T)=Z_0(T-1)$. The single-field bosonic Lagrangian obtained in this procedure (see the Appendix~\ref{appendix}) is
\be 
\mathcal{L}=-3\frac{|\partial_\mu T|^2}{(T+\overline{T})^2}-\frac{|Z(T)|^2}{(T+\overline{T})^2}\,.
\label{NOSCALELANG}
\ee
This Lagrangian will be matched to those derived from metric-affine models. Similar scenarios may also be realized by fixing the modulus $T$ and allowing $S$ to play the role of the inflaton~\cite{Ellis:2013xoa,Ellis:2013nxa}. In this setup, the Starobinsky model is obtained by choosing the superpotential to take the standard Wess-Zumino form~\cite{Wess:1974tw}, namely $W(S)=M(S^2/2-\lambda S^3/3\sqrt{3})$. In the following, however, we focus on the first approach, where $S\rightarrow 0$ and Re$(T)$ plays the role of the inflaton.

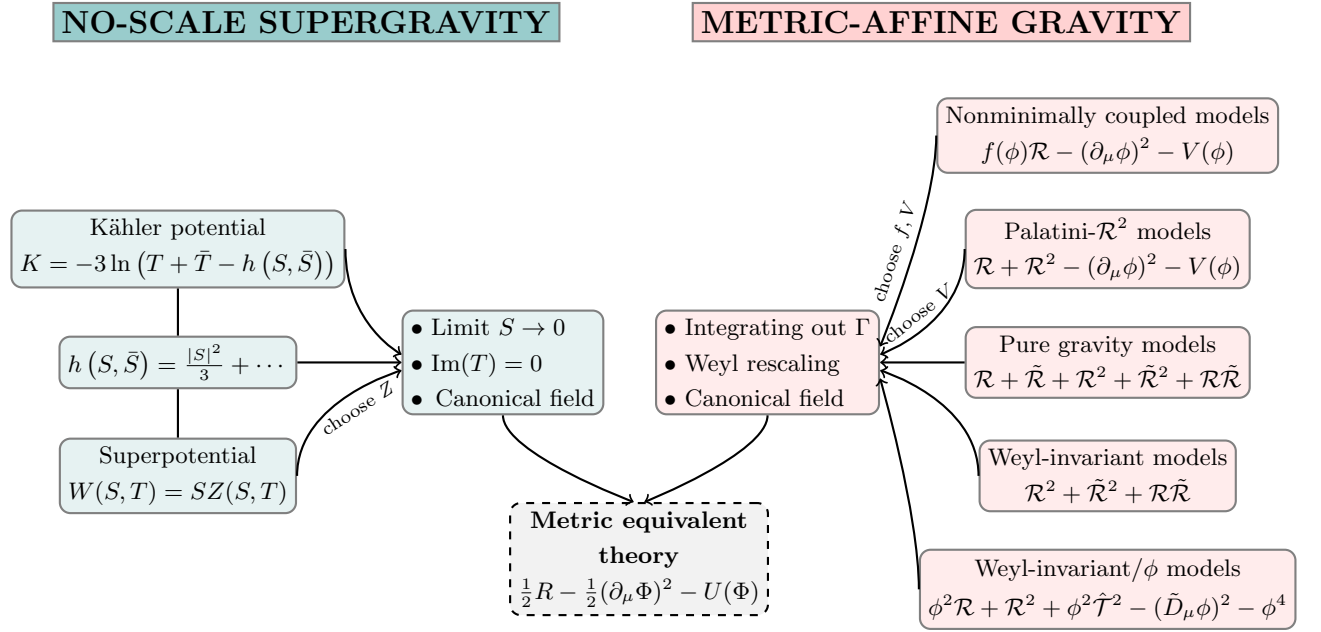
\begin{figure*}[t!]
\centering
\begin{tikzpicture}[auto, scale=1, every node/.style={scale=1}]

\tikzstyle{sugracolor}=[rectangle,draw=black!50,fill=teal!40,thick]
\tikzstyle{sucolors}=[rectangle,rounded corners,draw=black!50,fill=teal!10,thick]

\tikzstyle{magcolor}=[rectangle,draw=black!50,fill=pink!70,thick]
\tikzstyle{mcolors}=[rectangle,rounded corners,draw=black!50,fill=pink!30,thick]

\tikzstyle{comcolor}=[rectangle,rounded corners,draw=black!100,dashed,fill=gray!10,thick]

\node[align=center] (sugra) at (-0.2,0.7) [sugracolor] {\textbf{\large{NO-SCALE SUPERGRAVITY}}};

\node[align=center] (mag) at (8.2,0.7) [magcolor] {\textbf{\large{METRIC-AFFINE GRAVITY}}};

\node[align=center] (SUGRA0) at (-1.9,-2.3) [sucolors] {K{\"a}hler potential\\[1mm]
$K=-3\ln\left(T+\bar{T}-h\left(S,\bar{S}\right)\right)$};
\node[align=center] (SUGRA1) at (-1.9,-3.8) [sucolors]{ $h\left(S,\bar{S}\right)= \frac{|S|^2}{3}+\cdots$};
\node[align=center] (SUGRA2) at (-1.9,-5.3) [sucolors] {Superpotential\\[1mm]
$W(S,T) = S Z(S,T)$};

\node[align=left] (SUGRA3) at (2.4,-3.8) [sucolors] { $\bullet$ Limit $S\rightarrow 0$\\[1mm] $\bullet$ ${\rm Im}(T)=0$\\[1mm]
$\bullet\, $ Canonical field};

\node[align=center] (MAG0) at (10.4,-0.8) [mcolors] {Nonminimally  coupled models \\[1mm]$f(\phi)\mathcal{R}-(\partial_\mu\phi)^2-V(\phi)$};
\node[align=center] (MAG1) at (10.4,-2.3) [mcolors] {Palatini-$\mathcal{R}^2$ models \\[1mm] $\mathcal{R}+\mathcal{R}^2-(\partial_\mu\phi)^2-V(\phi)$};
\node[align=center] (MAG2) at (10.4,-3.8) [mcolors] {Pure gravity models \\[1mm]$\mathcal{R}+\tilde{\mathcal{R}}+\mathcal{R}^2+\tilde{\mathcal{R}}^2+\mathcal{R}\tilde{\mathcal{R}}$};
\node[align=center] (MAG2b) at (10.4,-5.3) [mcolors] {Weyl-invariant models \\[1mm]$\mathcal{R}^2+\tilde{\mathcal{R}}^2+\mathcal{R}\tilde{\mathcal{R}}$};
\node[align=center] (MAG2c) at (10.4,-6.8) [mcolors] {Weyl-invariant/$\phi$ models\\[1mm]$\phi^2\mathcal{R}+\mathcal{R}^2+\phi^2\hat{\mathcal{T}}^2-(\tilde{D}_\mu\phi)^2-\phi^4$};
\node[align=left] (MAG3) at (5.9,-3.8) [mcolors] {$\bullet$ Integrating out $\Gamma$\\[1mm] $\bullet$ Weyl rescaling\\[1mm] $\bullet$ Canonical field};

\node[align=center] (common0) at (4.2,-6.4) [comcolor] {\textbf{Metric equivalent}\\[1mm] \textbf{theory} \\[1mm]  $\frac{1}{2}R-\frac{1}{2}(\partial_\mu\Phi)^2-U(\Phi)$};

\draw[thick,black,-] (SUGRA0.south) -- (SUGRA1.north);
\draw[thick,black,-] (SUGRA1.south) -- (SUGRA2.north);

\draw[thick,black,->] (SUGRA0.east) .. controls +(0,-1) and +(0,0.) .. ([yshift=3pt]SUGRA3.west);
\draw[thick,black,->] (SUGRA1.east) .. controls +(0,0) and +(0,0.) .. (SUGRA3.west);
\draw[thick,black,->] (SUGRA2.east) .. controls +(0,1) and +(0,0.) .. ([yshift=-3pt]SUGRA3.west)
node[draw=none,fill=none,font=\scriptsize,midway,below,sloped] {choose $Z$};

\draw[thick,black,->] (MAG0.west) .. controls +(0,-1) and +(0,0.) .. ([yshift=6pt]MAG3.east)
node[draw=none,fill=none,font=\scriptsize,midway,above,sloped,xshift=5pt,yshift=0pt] {choose $f,V$};

\draw[thick,black,->] (MAG1.west) .. controls +(0,-1) and +(0,0.) .. ([yshift=3pt]MAG3.east)
node[draw=none,fill=none,font=\scriptsize,midway,above,sloped,xshift=3pt,yshift=0pt] {choose $V$};

\draw[thick,black,->] (MAG2.west) .. controls +(0,0) and +(0,0.) .. ([yshift=0pt]MAG3.east);

\draw[thick,black,->] (MAG2b.west) .. controls +(0,1) and +(0,0.) .. ([yshift=-3pt]MAG3.east);

\draw[thick,black,->] (MAG2c.west) .. controls +(0,1) and +(0,0.) .. ([yshift=-6pt]MAG3.east);

\draw[thick,black,->] (SUGRA3.south) .. controls +(0,-0.5) and +(0,0.) .. ([xshift=-2pt]common0.north);
\draw[thick,black,->] (MAG3.south) .. controls +(0,-0.5) and +(0,0.) .. ([xshift=2pt]common0.north);

\end{tikzpicture}
\caption{Visualization of the ``bridge'' linking no-scale supergravity with metric-affine gravity models. For simplicity, constants and numerical coefficients are omitted in the metric-affine models shown, though they are retained in the main text.}
\label{fig:bridge}
\end{figure*}

It should be emphasized that the relation between the metric-affine and supergravity frameworks discussed in this work is not a strict correspondence but rather a formal analogy that provides intuition for constructing viable inflationary potentials within no-scale supergravity.

\section{Models}
\label{models}
In this section, we analyze popular metric-affine models derived from the general action~\eqref{GENLANG}, and identify suitable superpotentials of the form~\eqref{eq:super_pot} that can reproduce these models within the framework of no-scale supergravity. Figure~\ref{fig:bridge} illustrates the connection between metric-affine and no-scale models.

\subsection{Palatini-\texorpdfstring{$\mathcal{R}^2$}{+R\^{}2} models}
\label{modelA}

We start considering the simple case of the so-called Palatini-$R^2$ models defined by the Jordan frame (JF) action
\be
\label{eq:palR2}
{\cal{S}}=\int {\rm d}^4x \sqrt{-g}\left[\frac{1}{2}{\cal{R}}+\frac{\gamma}{4}{\cal{R}}^2-\frac{1}{2}(\partial_\mu\phi)^2-V(\phi)\right]\,.
\ee
Going through the steps described above, we arrive at the equivalent Einstein-frame (EF) action
\be 
{\cal{S}}=\int {\rm d}^4x\sqrt{-g}\left[\frac{R}{2}-\frac{1}{2}\frac{(\partial_\mu\phi)^2}{(1+4\gamma V(\phi))}-\frac{V(\phi)}{(1+4\gamma V(\phi))}\right]\,,{\label{EINSTEIN-0}}
\ee
which has been shown to exhibit a by default-inflationary behavior, with 
an EF potential plateau in the regime $V(\phi)\gg 1/(4\gamma)$~\cite{Meng:2004yf,Enckell:2018hmo,Antoniadis:2018ywb,Antoniadis:2018yfq,Tenkanen:2019jiq,Gialamas:2019nly}. We have neglected the arising $\mathcal{O}((\partial_\mu \phi)^4)$ term.
The action~\eqref{EINSTEIN-0} can in principle be set in canonical form 
in terms of the JF potential $V(\phi)$, by defining  $U(\phi)\equiv V(\phi)/(1+4\gamma V(\phi))$, and introducing a canonical field $\Phi$ as ${\rm d}\Phi={\rm d}\phi/\sqrt{1+4\gamma V(\phi)}$.

As a simple case that can be treated analytically we consider the case of a quadratic JF potential $V(\phi)=m^2\phi^2/2$
for which the canonical field $\Phi$ comes out to be $\Phi=\sinh^{-1}(m\sqrt{2\gamma}\,\phi)/(m\sqrt{2\gamma})$.
The EF potential, expressed in terms of the canonical field, reads
\be 
U(\Phi)=\frac{1}{4\gamma}\tanh^2(m\sqrt{2\gamma}\,\Phi)\,.
\ee
It is now straightforward to compare the Palatini-$R^2$ model in its canonical form to the general no-scale Lagrangian~\eqref{NOSCALELANG} and deduce the corresponding superpotential function $Z(T)$. For real fields, i.e. Im$(T)=0$, we have
\be 
\Phi=\sqrt{\frac{3}{2}}\ln T \quad \text{and} \quad U(\Phi(T))=\frac{1}{4\gamma}\left(\frac{T^{\nu}-1}{T^{\nu}+1}\right)^2\,,
\ee
where $\nu=2m\sqrt{3\gamma}$. From this we obtain
\be
\label{zeta_R2}
Z(T)=\pm\frac{T}{\sqrt{\gamma}}\left(\frac{T^{\nu}-1}{T^{\nu}+1}\right)\,.
\ee
The choice of the superpotential~\eqref{eq:super_pot}, with
$Z(T)$ given by~\eqref{zeta_R2}, and assuming Im$(T) = 0$ and $S \rightarrow 0$, reduces to the
Palatini-$R^2$ model~\eqref{eq:palR2} with a quadratic potential.

\subsection{Nonminimally coupled models}
\label{modelB}

The general features of the above correspondence between the metric-affine ${\cal{R}}^2$ models and appropriate no-scale supergravity hold true also in the case that nonminimal couplings of the scalars to curvature are present~\cite{Koivisto:2005yc,Bauer:2008zj,Rasanen:2017ivk,Racioppi:2017spw,Markkanen:2017tun,Jarv:2017azx,Fu:2017iqg,Racioppi:2018zoy}
\be 
{\cal{S}}=\int {\rm d}^4x \sqrt{-g}\left[\frac{1}{2}f(\phi){\cal{R}}-\frac{1}{2}(\partial_\mu\phi)^2-V(\phi)\right]\,.
\ee
The equivalent metric action in the EF is
\be {\cal{S}}=\int{\rm d}^4x\sqrt{-g}\left[\frac{1}{2}R-\frac{1}{2}\frac{(\partial_\mu\phi)^2}{f(\phi)}-\frac{V(\phi)}{f^2(\phi)}\right]\,.\ee
To illustrate the correspondence we focus on the particular case where the coupling function is given by $f(\phi)=1+\xi\phi^2$ and the potential takes the monomial form $V(\phi)=\lambda_n\phi^n$. 
The corresponding canonical field is $\Phi = \sinh^{-1}(\sqrt{\xi}\phi)/\sqrt{\xi}$ and the EF scalar potential turns out to be
\begin{equation}
U(\Phi)=\frac{\lambda_n}{\xi^{n/2}}\frac{\sinh^n(\sqrt{\xi} \Phi)}{\cosh^4(\sqrt{\xi}\Phi)}\,.
\end{equation}
Comparing this with the no-scale potential in~\eqref{NOSCALELANG} we obtain
\begin{equation}
    Z(T)=\pm Z_0\frac{T^{1+(1-n/4)\nu}\left(T^{\nu}-1\right)^{n/2}}{\left(T^{ \nu}+1\right)^2}\,,
\end{equation}
with $Z_0=8\sqrt{\lambda_n}(4\xi)^{-n/4}$ and $\nu=\sqrt{6\xi}$. 
In the special case of a Higgs-like quartic potential~\cite{Bezrukov:2007ep,Bezrukov:2010jz,Shaposhnikov:2020fdv}, i.e. $n=4$, we obtain
\begin{equation}
    Z(T)=\pm Z_0T^{}\frac{(T^{\nu}-1)^{2}}{(T^{\nu}+1)^2}\,.
\end{equation}

\subsection{Pure gravitational action}
\label{modelC}

Let us consider now the purely gravitational general quadratic metric-affine action
\be
\label{eq:action_holst}
{\cal{S}}=\int {\rm d}^4x\sqrt{-g}\left[\frac{1}{2}{\cal{R}}+\frac{\beta}{2}\tilde{\cal{R}}+\frac{\gamma}{4}{\cal{R}}^2+\frac{\delta}{4}\tilde{\cal{R}}^2+\frac{\epsilon}{2}{\cal{R}}\tilde{\cal{R}}\right]\,,
\ee
where, in addition to the Ricci scalar, we have also included the other possible curvature scalar, namely the {\textit{Holst}} invariant
$\tilde{\cal{R}}$. This purely gravitational model has been shown~\cite{Pradisi:2022nmh} to contain, in addition to the standard graviton, a pseudoscalar particle associated with the Holst term.
The corresponding EF action for this model can once again be expressed in terms of a canonically normalized field and a potential given by~\cite{Gialamas:2022xtt}
\begin{equation}
\label{eq:Holst_pot}
U(\Phi)=\frac{\left(\sinh(\sqrt{2/3}\,\Phi)-2\beta\right)^2}{16(\beta^2\gamma+\delta-2\beta\epsilon)}\,.
\end{equation}
Given that $T=e^{\sqrt{\frac{2}{3}}\Phi}$, we can immediately read off from ({\ref{NOSCALELANG}}) that
\be
\label{eq:zeta_holst}
Z(T)=\pm\frac{Z_0}{4}(T^2-1-4\beta T)\,,
\ee
where $Z_0 =1/ \sqrt{\beta^2\gamma+\delta-2\beta\epsilon} $. Actually, rescaling $T\rightarrow -T/4\beta$ and taking the limit $\beta \rightarrow\infty$ we obtain the standard Starobinsky model~\cite{Racioppi:2024pno}.
Actions of the form~\eqref{eq:action_holst}, which include both minimal and nonminimal couplings between matter and curvature, have been extensively studied in the context of inflationary cosmology~\cite{Langvik:2020nrs,Shaposhnikov:2020gts,Salvio:2022suk,Gialamas:2022xtt,DiMarco:2023ncs,He:2024wqv,Racioppi:2024zva,Racioppi:2024pno,Gialamas:2024uar,Katsoulas:2025mcu}. A notable feature of the potential~\eqref{eq:Holst_pot}, derived from the purely gravitational action~\eqref{eq:action_holst}, is its flattening, which results from the inclusion of the parity-violating term linear in $\tilde{\mathcal{R}}$. This flattening is crucial for realizing a viable inflationary scenario, as it ensures consistency with observational data. The remaining terms are subdominant, mainly affecting the energy scale of inflation.

Apart from the fact that the corresponding no-scale model has a rather simple structure, the main attractive feature of this model is that it is purely gravitational.

\subsection{Weyl-invariant Einstein-Cartan models}
\label{modelD}

As a next step, we continue our consideration of models within the Einstein–Cartan framework, which are invariant under local \textit{Weyl rescalings}. Having restricted ourselves to at most quadratic curvature terms in the action for the reasons stated in the introduction, the only possible purely gravitational Weyl-invariant action is~\cite{Karananas:2024xja,Gialamas:2024iyu,Karananas:2025xcv,Gialamas:2025ciw}
\be 
\mathcal{S}=\int {\rm d}^4x\sqrt{-g}\left[\frac{\gamma}{4}\mathcal{R}^2+\frac{\delta}{4}\mathcal{\Tilde{R}}^2 +\frac{\epsilon}{2}\mathcal{R} \mathcal{\Tilde{R}}\right]\,.
\ee
Introducing auxiliary fields $\chi$ and $\zeta$, this action can be rewritten in the equivalent form
\begin{align}
\mathcal{S}=\int {\rm d}^4x\sqrt{-g}\bigg[&\frac{1}{2}(\gamma\chi+\epsilon\zeta)\mathcal{R}+\frac{1}{2}(\delta \zeta+\epsilon\chi)\mathcal{\tilde{R}} \nonumber
\\
&-\frac{\gamma}{4}\chi^2-\frac{\delta}{4}\zeta^2-\frac{\epsilon}{2}\chi\zeta  \bigg]\,.
\end{align}
Weyl invariance allows us to fix the gauge by choosing $\chi=1/\gamma$. Introducing the expressions of ${\cal{R}}$ and $\tilde{\cal{R}}$ in terms of metric quantities and the torsion and integrating out the latter, we arrive at an equivalent EF metric action, featuring an additional pseudoscalar degree of freedom. A suitable field redefinition reduces the action into a model of a canonically normalized scalar field $\Phi$ with a potential
\begin{equation}
U(\Phi) =\frac{1}{4\gamma}+\frac{\gamma}{16}\frac{\left(\sinh(\sqrt{2/3}\Phi)-\frac{2\epsilon}{\gamma}\right)^2}{(\gamma\delta-\epsilon^2)}\,.   
\end{equation}
This potential can be matched to a no-scale superpotential function
\begin{equation}
Z^2(T)=\frac{T^2}{\gamma}+\frac{\gamma}{16(\gamma\delta-\epsilon^2)}\left(T^2-1-\frac{4\epsilon}{\gamma}T\right)^2\,,
\end{equation}
which in the limit $\gamma\gg1$ reduces to
\begin{equation}
    Z(T)\simeq\pm \frac{1}{4\sqrt{\delta}}(T^2-1)\,.
\end{equation}

\subsection{Weyl-invariant models coupled to matter}
\label{modelE}

Local Weyl-invariance within the Einstein-Cartan framework consists in promoting the Poincar\'e group to a local gauge symmetry. Within this framework, curvature and torsion are associated with local translations and Lorentz transformations, respectively. The most general action consistent with Weyl invariance and restricted to terms involving at most two derivatives of the fields is given by~\cite{Karananas:2021gco,Karananas:2024xja,Gialamas:2024iyu}
\begin{align}
\mathcal{S} = \int {\rm d}^4x \sqrt{-g}&\left[\frac{\xi\phi^2}{2}\mathcal{R} +\frac{\gamma}{4}\mathcal{R}^2 -\frac12(\tilde{D}_\mu\phi)^2 \right. \nonumber
\\
& \hspace{0.2cm}\left.-\frac{\lambda}{4}\phi^4 +\mathcal{C}\phi^2\hat{\mathcal{T}}^2\right]\,,
\end{align}
where the Weyl covariant derivative is defined as $\tilde{D}_\mu\phi \equiv \partial_\mu\phi +\mathcal{T}_\mu\phi/3$. Note that the torsion vector ${\cal{T}}_{ \mu}$ is identified with the gauge vector field of the local Weyl invariance. Terms involving the Holst invariant, such as, $\sim\phi^2\tilde{\mathcal{R}}$ and $\sim\tilde{\mathcal{R}}^2 $ are also allowed. Nevertheless, they have been omitted. Their presence introduces an additional pseudoscalar degree of freedom, as in the case of the previously analyzed purely gravitational example, promoting the model into a two-field model and complicating the correspondence with no-scale supergravity models\footnote{ We also omit terms involving the tensorial component $\tau_{\mu\nu\rho}$, specifically, $\sim \phi^2 \tau_{\mu\nu\rho} \tau^{\mu\nu\rho}$ and $\phi^2\epsilon^{\mu\nu\rho\sigma} \tau_{\lambda\mu\nu}\tensor{\tau}{^\lambda_\rho_\sigma}$, since it enters the action quadratically, leading to $\tau_{\mu\nu\rho}=0$.}.
Varying the action with respect to the the torsion we obtain a set of algebraic equations, the solution of which is substituted back into the action. The resulting EF action is
\begin{align}
\label{eq:weylEF}
\mathcal{S} = \int {\rm d}^4x \sqrt{-g}\bigg[&\frac12R -\frac{3(\partial_\mu\phi)^2}{(1+\xi\phi^2)^2(6+(1+6\xi)\phi^2)} \nonumber
\\
&-\frac{1+\gamma\lambda\phi^4}{4\gamma(1+\xi\phi^2)^2}\bigg]\,.   
\end{align}
Upon applying the appropriate field redefinition, $\Phi = \sqrt{6}\tanh^{-1}[\phi/\sqrt{6+(1+6\xi)\phi^2}]$, the action is transformed into that of a canonical scalar field $\Phi$ with a potential given by
\begin{equation}
    U(\Phi) = 9\lambda\sinh^4\left[\Phi/\sqrt{6}\right]+\frac{1}{4\gamma} \left(1-6\xi\sinh^2\left[\Phi/\sqrt{6}\right] \right)^2.
\end{equation}
This potential can be matched to that of a no-scale supergravity model, with the function
\begin{equation}
\label{eq:zeta_Weyl_field}
   Z^2(T)=\frac{1}{\gamma}\left(T^2+\frac{9}{4}(\lambda\gamma+\xi^2)(T-1)^4-3\xi T(T-1)^2\right)\,,
\end{equation}
which for $\lambda\gamma\ll\xi^2$ reduces to
\begin{equation}
    Z(T)\simeq \pm\frac{ 1}{\sqrt{\gamma}}\left(\frac{3\xi}{2}(T-1)^2-T\right)\,.
\end{equation}

\newpage
\section{Conclusions}
\label{conclusions}


 In this paper, we studied a class of no-scale supergravity models whose structure was motivated by certain metric-affine gravitational models known to yield successful inflationary dynamics. 
While no direct or fundamental correspondence between the two frameworks is established, the analogy provides a useful perspective: the noncanonical kinetic terms and scalar potentials that arise naturally in the metric-affine context can be effectively realized within no-scale supergravity through suitably chosen superpotentials. 
This observation broadens the scope of viable no-scale constructions and illustrates how insights from alternative gravitational formulations can guide supergravity model building. 
Our results demonstrate that the no-scale framework can accommodate rich inflationary dynamics consistent with current observational constraints.

\section*{ACKNOWLEDGMENTS}
IDG thanks Antonio Racioppi for useful discussions. The work of IDG was supported by the Estonian Research Council grants MOB3JD1202, RVTT3, RVTT7 and by the CoE program TK202 ``Foundations of the Universe'’.

\appendix
\section{NO-SCALE SUPERGRAVITY SYLLABUS}
\label{appendix}

The no-scale supergravity framework in its simplest $SU(2,1)/SU(2)\times U(1)$ form, is defined in terms of a K{\"a}hler potential
\be {\cal{K}}\,=\,-3\ln\left(T+\bar{T}-h(S,\bar{S})\,\right)\,,{\label{NOSCAL}}\ee
where $T$ and $S$ are two chiral superfields. The scalar potential of the model, resulting from the K{\"a}hler potential ${\cal{K}}$ and a superpotential $W(T,S)$ is, in terms of the function $G={\cal{K}}+\ln|W|^2$, 
\be V=e^{G}\left(G_{\bar{i}}\left(K^{-1}\right)_{\bar{i}j}G_j-3\right)\,,
\ee
with
\be 
G_i=\frac{\partial G}{\partial\phi_i}\,, \quad K_{i\bar{j}} =\frac{\partial^2K}{\partial\phi_i\partial\bar{\phi}_{\bar{j}}}\,.
\ee
The kinetic Lagrangian of the bosonic $(T,S)$-sector is
\be 
{\cal{L}}_{kin}=-{\cal{K}}_{i\bar{j}}(\partial_\mu\phi_i)(\partial^\mu\bar{\phi}_{\bar{j}})\,.
\ee
The overall bosonic Lagrangian resulting from the K{\"a}hler potential~\eqref{NOSCAL} and a general superpotential $W(S,T)$ is
\begin{widetext}
\begin{equation}
\begin{aligned}
{\cal{L}}= &-\frac{3|\partial_\mu T|^2}{(T+\bar{T}-h)^2}-\frac{3|\partial_\mu S|^2}{(T+\bar{T}-h)^2}\left((T+\bar{T}-h)h_{S\bar{S}}+|h_S|^2\right)  +\frac{3\left(h_S\partial_\mu S \partial^\mu\bar{T}+c.c.\right)}{(T+\bar{T}-h)^2} 
\\
& +\frac{(W_T\bar{W}+c.c.)}{(T+\bar{T}-h)^2}-\frac{|W_S|^2}{3h_{S\bar{S}}(T+\bar{T}-h)^2}
 -\frac{|W_T|^2}{3(T+\bar{T}-h)}-\frac{\left(|W_T|^2|h_S|^2+\bar{W}_{\bar{T}}W_Sh_{\bar{S}}+W_T\bar{W}_{\bar{S}}h_S\right)}{3h_{S\bar{S}}(T+\bar{T}-h)^2}\,,
\end{aligned}
\end{equation}    
\end{widetext}
where $W_T$ and $W_S$ denote the partial derivatives of the superpotential with respect to $T$ and $S$ respectively, and $h_S,\,h_{S\bar{S}}$ are the corresponding derivatives of $h$.
In general, the function $h(S,\bar{S})$ is a function with $h(0)=0$ and $h_{S,\bar{S}}(0)>0$, while for the $SU(2,1)/SU(2)\times U(1)$ model we take $h(S,\bar{S})=|S|^2/3+\cdots$, where the dots signify higher order terms, required for stability purposes.

\bibliography{references}

\end{document}